\documentclass[pra,a4paper,twocolumn,showpacs,superscriptaddress]{revtex4-1}
\usepackage{amsmath}
\usepackage{amsfonts}
\usepackage{graphicx}
\usepackage{longtable}
\begin{document}

\title{Open system quantum dynamics with correlated initial states, \\  not completely positive maps and non-Markovianity} 
\author{A. R. Usha Devi}
\email{arutth@rediffmail.com}
\affiliation{Department of Physics, Bangalore University, 
Bangalore-560 056, India}
\affiliation{Inspire Institute Inc., Alexandria, Virginia, 22303, USA.}
\author{A. K. Rajagopal} 
\affiliation{Inspire Institute Inc., Alexandria, Virginia, 22303, USA.}
\author{Sudha} 
\affiliation{Department of Physics, Kuvempu University, Shankaraghatta, Shimoga-577 451, India.}
\affiliation{DAMTP, Centre for Mathematical Sciences, Wilberforce Road, Cambridge, CB3 0WA, UK.}
\date{\today}

\begin{abstract} 
  Dynamical $A$ and $B$ maps have been employed extensively by Sudarshan and co-workers to investigate open system evolution of quantum systems. 
  A canonical structure of the $A$-map is introduced here. It is shown that this canonical  $A$-map  enables us to investigate if the dynamics is  completely positive (CP) or non-completely positive (NCP) in an elegant way  and hence, it subsumes the basic results on open system dynamics. Identifying memory effects in open system evolution is gaining increasing importance recently and here, a criterion of non-Markovianity, based on the relative entropy of the dynamical state is proposed. The {\em relative entropy difference} of the dynamical system  serves as a complementary characterization -- though not related directly  -- to the {\em fidelity difference} criterion proposed recently. Three typical examples of open system evolution of a qubit, prepared initially in a correlated state with another  qubit (environment), and evolving jointly under a specific unitary dynamics -- which corresponds to a NCP dynamical map --  are investigated by employing both the relative entropy difference and fidelity difference tests of non-Markovianity. The two qubit initial states are chosen to be (i) a pure entangled state, (ii) the  Werner state, which exemplifies both entangled and separable states of qubits, depending on  a real parameter, and (3) a separable mixed state.  Both the relative entropy and fidelity  criteria offer a nice display of how non-Markovianity manifests  in all the three  examples. 
\end{abstract}
\maketitle
\section{Introduction} 
It is well-known~\cite{Breuer} that when a quantum system -- chosen initially to be in a tensor product state with its environmental degrees of freddom --  undergoes dynamical evolution, the final state of the system is related to the initial one via a completely positive (CP) dynamical map.  Kraus decomposition~\cite{Kraus} of dynamics is guaranteed only when the map is completely positive. After its conceptual formulation~\cite{ECGS1,Jordan1} nearly five decades ago,  Sudarshan and coworkers~\cite{Jordan,Rosario,ECGS,Modi} have been investigating the quantum theory of open system evolution in terms of dynamical maps in the more general setting -- including not completely positive (NCP) maps. There has been growing interest~\cite{Pechukas,Alicki,Buzek,Terno,Lidar} in identifying the physical conditions under which open evolution of a quantum system {\em does not} ensure a CP dynamical map.  Jordan et. al.~\cite{Jordan} studied an open system unitary evolution, where the system and the environment may be in an initially entangled state and showed that the resulting dynamical map is not always completely positive. Rodr{\' i}guez and Sudarshan~\cite{ECGS} analyzed the general characteristics of dynamical maps in open quantum system evolutions, taking into account initial correlations of system with environment. Extending the result of Ref.~\cite{Rosario}, Shabani and Lidar~\cite{Lidar} showed recently that CP maps are guaranteed for quantum dynamical processes, if and only if the initial system-environment state belongs to a class of separable states with vanishing quantum discord~\cite{Zurek}.    All these investigations point towards the important role of the initial state in open system evolution and more recently, Modi and Sudarshan~\cite{Modi} outlined the effects of preparation of initial state in quantum process tomography. 

 The Markov approximation, where the correlation time between the system and environment is considered to be infinitesimally small -- so that the dynamical map does not carry any memory effects -- leads to a much simplified picture of open system dynamics. The mathematical theory of Markovian dynamics is built around a CP  map -- originating from the unitary evolution of an initially uncorrelated system-environment state and generating  a  dynamical semigroup (to take into account the memorylessness). This results in the  Markovian dynamical equation -- known as the Lindblad-Gorini-Kossakowski-Sudarshan  (LGKS)~\cite{Lindblad,GKS} master equation -- for the time evolution of the system density matrix.  However, memory effects are prevalent in many physical situations of interest. Formulation of  CP non-Markovian processes --  where the dynamical evolution depends on the history of the system-environment correlation -- has attracted significant attention~\cite{ECGS, Daffer, Breuer2, Kossakowski}. Various manifestations of non-Markovianity have been investigated recently~\cite{Cirac, B3,Angel,AKRU}, based on the departure from strict Markovian behavior (where the dynamical map is a one-parameter continuous, memoryless, completely positive semigroup). However, revelation of non-Markovian features under NCP maps has not been studied so far. In the present paper, we focus on two basic issues: (1) A canonical structure of the dynamical map to ellucidate  CP or NCP nature  of dynamics and (2) operational signatures of non-Markovianity. We consider a particular unitary time evolution -- which has been shown~\cite{Jordan} to correspond to a NCP map  --  and examine the non-Markovianity of the open system  dynamics of single qubit systems, resulting from the  joint evolution of   different types of initially correlated two qubit states. We propose a criterion, based on the relative entropy of the quantum states to verify Markovianity/non-Markovianity of the dynamical process. We compare the results with the fidelity difference~\cite{AKRU} test of non-Markovianity proposed recently by some of us~\cite{AKRU}.   
 
We first review the properties of $A$ and $B$ maps~\cite{ECGS1, Rosario}, introduced in the general theory of open system dynamics in Sec.~II. Here, we present a new canonical structure of the $A$ map, which reveals its equivalence with that of the $B$ map. We employ this canonical map to establish CP or NCP nature of the dynamics.  In Sec.~III we explore a specific example of open system unitary evolution~\cite{Jordan} of a single qubit (system), which is initially prepared in a correlated state with another qubit (environment), leading to a NCP dynamical map. This map serves as a template for studying the different examples of intially correlated two qubit states considered here.  We construct a characterization of non-Markovianity in Sec.~IV, based on relative entropy of the evolving quantum states. We also discuss the fidelity difference criterion~\cite{AKRU} of non-Markovianity proposed earlier so as to be self-contained and mutually concordant.  We then proceed to illustrate non-Markoivan behaviour of the NCP dynamical map of Sec.~III, with the help of  three different choices of density matrices  prepared initially in entangled pure, mixed, and separable states of the two qubits. By employing the relative entropy criterion, as well as the  fidelity difference test, we show that the dynamics in all the three different cases exhibits non-Markovianity.  Sec. IV is devoted to a brief summary of our results. 

\section{Review on the general properties of $A$, $B$ dynamical maps and their equivalence} 

A general open system dynamics  relates the elements $\left[\rho(0)\right]_{rs}$ of the initial system density matrix with  $\left[\rho(t)\right]_{r's'}$  at instant $t$ via a linear map~\cite{ECGS1,Rosario}, 
\begin{equation}
\label{defA}
\left[\rho(t)\right]_{r's'}=\sum_{r,s=1}^{n}\, A_{r's';rs}(t)\, \left[\rho(0)\right]_{rs}, \ \ \  r',s'=1,2,\ldots, n.
\end{equation} 
Imposing that the $A$ map ensures (i) the preservation of hermiticity  i.e., $\left[\rho(t)\right]_{r's'}=\left[\rho(t)\right]^*_{s'r'}$,  and  (ii) unit trace condition  i.e., ${\rm Tr}[\rho(t)]=1$, the following restrictions on the elements of $A$ are realized~\cite{ECGS1,Rosario}: 
\begin{eqnarray}
\label{A1}
A_{s'r';sr}&=&A^{*}_{r's';rs}  \\
\label{A2}
\sum_{r'}\,A_{r'r';rs}&=&\delta_{r,s}.
\end{eqnarray}  
In order to bring out the properties (\ref{A1}), (\ref{A2}) in a transparent manner, it has been found convenient to define a realigned matrix $B$  ~\cite{ECGS1,Rosario}:
\begin{equation}
\label{abdef}
B_{r'r;s's}=A_{r's';rs}. 
\end{equation}  
The hermiticity property (\ref{A1}) leads to the condition $B_{s's;r'r}=B^*_{r'r;s's}$, i.e., the dynamical map $B$ itself is hermitian -- which is exploited further to identify the general features of dynamics~\cite{ECGS1,Rosario}.   
Here, we present an alternate version by casting the $A$-map in its canonical form, which is shown to capture all the dynamical features in an unique way.  To accomplish this, we start by  considering an orthonormal set $\{T_\alpha, \alpha=1,2,\ldots, n^2\}$ of $n\times n$ basis matrices , satisfying 
\begin{equation} 
{\rm Tr}[T_\alpha^\dag\, T_\beta]=\delta_{\alpha,\beta},
\end{equation} 
so that we can express the $n^2\times n^2$ matrix $A$ as,  
\begin{eqnarray}
\label{a1}
A&=&\sum_{\alpha\beta} {\cal A}_{\alpha\beta}\, T_\alpha\otimes T_\beta^*, \\ 
 {\cal A}_{\alpha\beta}&=&{\rm Tr}[A(T^\dag_\alpha\otimes T^T_\beta)]. 
\end{eqnarray} 
Clearly, we have, 
\begin{equation}
\label{a2}
A_{r's';rs}=\sum_{\alpha,\beta=1}^{n^2}\, {\cal A}_{\alpha\beta}\, [T_\alpha]_{r'r}\, [T^*_\beta]_{s's}.   
\end{equation} 
The hermiticity preservation condition (\ref{A1}) implies that 
\begin{equation}
{\cal A}_{\alpha\beta}={\cal A}^*_{\beta\alpha} 
\end{equation} 
i.e., the coefficients ${\cal A}_{\alpha\beta}$ form a $n^2\times n^2$ hermitian matrix ${\cal A}$. Denoting  ${\cal U}$ as the matrix  diagonalizing ${\cal A}$ and  $\{\lambda_\mu\}$,  the real eigenvalues of ${\cal A}$,  so that  $\sum_{\alpha,\beta} {\cal U}_{\mu \alpha}\, 
{\cal A}_{\alpha\beta} {\cal U}^*_{\mu\beta}= \lambda_\mu$,   we finally obtain the canonical structure of the $A$-map:
\begin{eqnarray}
\label{canA}
A&=&\sum_{\alpha,\beta,\mu}\,\lambda_\mu\, {\cal U}_{\mu \beta}{\cal U}^*_{\mu\alpha}\, T_\alpha\otimes T_\beta^* \nonumber \\
&=& \sum_\mu\, \lambda_\mu\, {\cal C}_\mu\otimes {\cal C}_\mu^*,
\end{eqnarray} 
where $ {\cal C}_\mu=\sum_\alpha\, {\cal U}^*_{\mu\alpha}\, T_\alpha$. With the help of the canonical form, the matrix elements of $A$ are explicitly given by,    
\begin{equation}
\label{canA2}
 A_{r's';rs}=\sum_\mu\, \lambda_\mu\, [{\cal C}_\mu]_{r'r}\,[{\cal C}_\mu^*]_{s's}.
\end{equation}
Substituting (\ref{canA2}) in (\ref{defA}) and simplifying, we obtain the following elegant form for the action of the $A$-map on the initial density matrix $\rho(0)$:   
\begin{equation}
\label{ele}
\rho(t)=\sum_{\mu}\lambda_{\mu}\, {\cal C}_{\mu}\, \rho(0)\,{\cal C}^\dag_{\mu}. 
\end{equation}   
The trace preservation condition (\ref{A2}) can be readily expressed as $\sum_{\mu}\,\lambda_\mu\, {\cal C}^\dag_{\mu}\, {\cal C}_{\mu}=I$ (where $I$ denotes  the $n\times n$ identity matrix).   The dynamical map is CP, when all the eigenvalues $\lambda_{\mu}$ are non-negative, whereas it is NCP if at least one of them is negative~\cite{Choi}.

From (\ref{abdef}) and (\ref{canA2}) we obtain, 
$B_{r'r;s's}=  \sum_\mu\, \lambda_\mu\, [{\cal C}_\mu]_{r'r}\, [{\cal C}^*_\mu]_{s's}$,
which evidently corresponds to the spectral decomposition of the $B$ matrix.  In other words, the  eigenvalues  of the matrix ${\cal A}$ and $B$ are identically same and   $[{\cal C}_\mu]_{r'r}$ (expressed as $n^2$ component column) correspond to the corresponding eigenvectors of $B$. 
In this paper we employ the canonical form of the $A$-map to elucidate the CP or NCP nature of the evolution. 

\section{An example of two qubit unitary dynamics}

Jordan et. al.~\cite{Jordan} studied a specific example of unitary dynamical evolution on two qubit states $U(t)=e^{-iHt/\hbar}$, governed by the Hamiltonian 
\begin{equation}
H=\frac{1}{2}\,\hbar\omega\, \sigma_{1z}\sigma_{2x},
\end{equation}
where $\sigma_{1\, i},\sigma_{2\, i}\, i=x,y,z$ respectively denote Pauli matrices of first and second qubits. 
The unitary tranformation matrix on the qubits is given explicitly (in the standard qubit basis
$\vert 0,0\rangle, \vert 0,1\rangle, \vert 1,0\rangle, \vert 1,1\rangle$) by, 
\begin{equation}
\label{ue}
 U(t)=\left(\begin{array}{cccc} \cos\left(\frac{\omega\, t}{2}\right) & -i\sin\left(\frac{\omega\, t}{2}\right) & 0 & 0\\ -i\sin
 \left(\frac{\omega\, t}{2}\right) & \cos\left(\frac{\omega\, t}{2}\right) & 0 & 0 \\ 
 0 & 0 & \cos\left(\frac{\omega\, t}{2}\right) & i\sin\left(\frac{\omega\, t}{2}\right) \\ 
 0 & 0 & i\sin\left(\frac{\omega\, t}{2}\right) & \cos\left(\frac{\omega\, t}{2}\right)
 \end{array}\right). 
 \end{equation} 
The positive definiteness of the dynamical state of the system qubit    --  evolving jointly with another environment qubit under the unitary time evolution (\ref{ue}) -- is necessarily preserved.
     
Time evolution of the expectation values  of the Pauli operators of the first qubit -- evaluated in the Heisenberg picture -- are given by, 
\begin{eqnarray}
\langle U^\dag(t) \sigma_{1x}U(t)\rangle&=&\langle\sigma_{1x}\rangle\,\cos(\omega\,t)-\langle\sigma_{1y}\sigma_{2x}\rangle\,\sin(\omega\,t) \nonumber \\
&=&\langle\sigma_{1x}\rangle\,\cos(\omega\,t)+a_1\,\sin(\omega\,t) \\
\langle U^\dag(t) \sigma_{1y}U(t)\rangle&=&\langle\sigma_{1y}\rangle\,\cos(\omega\,t)+\langle\sigma_{1x}\sigma_{2x}\rangle\,\sin(\omega\,t) \nonumber \\
&=&\langle\sigma_{1y}\rangle\,\cos(\omega\,t)+a_2\,\sin(\omega\,t) \\
\langle U^\dag(t) \sigma_{1z}U(t)\rangle&=&\langle\sigma_{1z}\rangle,
\end{eqnarray}  
where $\langle\sigma_{1x}\rangle,\ \langle\sigma_{1y}\rangle, \langle\sigma_{1z}\rangle$ are the expectation values  at $t=0$ and 
\begin{equation} 
\label{a1a2}
a_1=-\langle\sigma_{1y}\sigma_{2x}\rangle,\ a_2=\langle\sigma_{1x}\sigma_{2x}\rangle
\end{equation} 
 are considered to be the fixed initial state parameters describing the evolution of the first qubit~\cite{Jordan}. Correspondingly, the density matrix of the first qubit 
\begin{equation}
\rho_1(0)=\frac{1}{2}(I_1+\sigma_{1x}\, \langle\sigma_{1x}\rangle\,+\sigma_{1y}\, \langle\sigma_{1y}\rangle+ 
\sigma_{1z}\, \langle\sigma_{1z}\rangle)
\end{equation}    
is mapped to
\begin{eqnarray}
\label{ev}
\rho_1(t)&=&\frac{1}{2}\left[I_1+(a_1\, \sigma_{1x}+a_2\, \sigma_{1y})\, \sin(\omega t)+\sigma_{1x}\, \langle\sigma_{1x}\rangle\, \cos(\omega t) \right. \nonumber \\
 &&\left. +
\sigma_{1y}\, \langle\sigma_{1y}\rangle\, \cos(\omega t)
+\sigma_{1z}\, \langle\sigma_{1z}\rangle\right].  
\end{eqnarray}    
Eq. (\ref{ev}) in turn corresponds to~\cite{Jordan},  
\begin{eqnarray} 
\label{evop}
I_1'&=& I_1+(a_1\, \sigma_{1x}+a_2\, \sigma_{2x})\, \sin(\omega\,t),\nonumber \\ 
\sigma'_{1x}&=&\sigma_{1x}\, \cos(\omega\,t), \nonumber \\
\sigma'_{1y}&=&\sigma_{1y}\, \cos(\omega\,t), \nonumber \\
\sigma'_{1z}&=&\sigma_{1z}.
\end{eqnarray}
For fixed parameters $a_1,\, a_2$ characterizing the initial state, a linear dynamical $A$-map $Q \rightarrow Q'$ for all $2\times 2$ hermitian matrices -- consistent with the unitary evolution (\ref{ue}), is defined by (see Eq. (\ref{defA})), 
\begin{equation}
 Q'_{rs}=\sum_{r's'}\, A_{rs; r's'}\, Q_{r's'},\ \ r,s,r',s'=0,1, 
\end{equation}   
 where 
 \begin{equation}
 \label{Aex}
   A=\left(\begin{array}{cccc} 
  1 & 0 & 0 & 0 \\ \frac{1}{2}\,S\,a^* &  C & 0 & \frac{1}{2}\, S\,a^*\\  
 \frac{1}{2}\, S\, a & 0  & C & \frac{1}{2}\, S\, a \\ 
 0 & 0 & 0 & 1
 \end{array}\right), 
\end{equation}
with $a=a_1+i\,a_2;$ $C=\cos(\omega t)$, $S=\sin(\omega t)$  

Choosing $\left\{\frac{\sigma_{1\alpha}}{\sqrt{2}}\equiv \frac{I_1}{\sqrt{2}},\frac{\sigma_{1x}}{\sqrt{2}},\frac{\sigma_{1y}}{\sqrt{2}},\frac{\sigma_{1z}}{\sqrt{2}}\right\}$ as the orthonormal set of basis matrices, 
we expand the $A$ matrix (\ref{Aex}) as (see Eq.(\ref{a1})), 
\begin{eqnarray}
\label{a1'}
A&=&\frac{1}{2}\sum_{\alpha\beta} {\cal A}_{\alpha\beta}\, \sigma_\alpha\otimes \sigma_\beta^*, \\ 
 {\cal A}_{\alpha\beta}&=&\frac{1}{2}{\rm Tr}[A(\sigma_\alpha\otimes \sigma^*_\beta)]. 
\end{eqnarray} 
with the hermitian coefficient matrix ${\cal A}$ given by, 
\begin{eqnarray}
\label{calA}
{\cal A}&=&\frac{1}{2}{\rm Tr}[A(t)\sigma_\alpha\otimes\sigma_{\beta}^*] \nonumber \\
&=&\frac{1}{2}\, \left(\begin{array}{cccc} 
  2(1+C)\, & a_1\,S & a_2\,S & 0 \\ a_1\,S &  0 & 0 & i\,a_2\, S\\  
 a_2\,S& 0  & 0 & -i\,a_1\,S \\ 
 0 & -i\,a_2\, S & i\,a_1\, S & 2(1-C)
 \end{array}\right).
 \end{eqnarray}     
 The eigenvalues of ${\cal A}$ are given by~\cite{note}, 
 \begin{eqnarray}
 \label{lam4}
 \lambda_{1\pm}&=&\frac{1}{2}\left\{[1+\cos(\omega t)]\pm\sqrt{[1+\cos(\omega t)]^2+\vert a\vert^2\sin^2(\omega t)}\right\}\nonumber \\ 
 \lambda_{2\pm}&=&\frac{1}{2}\left\{[1-\cos(\omega t)]\pm\sqrt{[1-\cos(\omega t)]^2+\vert a\vert^2\sin^2(\omega t)}\right\}.\nonumber \\
 \end{eqnarray} 
 It may be seen that $\lambda_{1-},\lambda_{2-}$ assume negative values and thus, the dynamical map is NCP\cite{Jordan}(see however \cite{note2}). Nevertheless, as pointed out earlier, the positive definiteness of the dynamical single qubit  state, evolving jointly under the unitary dynamics (\ref{ue}) with another qubit (which are prepared in an intially correlated state) is always ensured. The $A$-map  (\ref{Aex}) serves as a general dynamical map for the different examples considered in Sec.~V.          
 
\section{Non-Markovian features}

We recall here that an open system CP dynamical map is Markovian if it forms a one parameter semigroup~\cite{Breuer}, which corresponds to 
\begin{equation}
\label{MarA}
A(t+\tau)=A(t)A(\tau), \ \ t,\tau\geq 0 
\end{equation} 
for the $A$-map (\ref{defA}). In other words,  when the underlying CP dynamics is Markovian, the $A$-map has an exponential structure $A=e^{t\,L}$, $L$ denoting the time-independent generator of the quantum dynamical semigroup~\cite{Lindblad,GKS}. One may verify directly whether the one-parameter semigroup criterion Eq.~(\ref{MarA}) is obeyed by checking if the $A$-map is exponential. However, it is advantageous to examine physical quantities, which are functions of the dynamical map, that describe the open system evolution of the physical states. In the following we consider relative entropy difference and fidelity difference to qualitatively capture the departure from the CP Markovian semigroup property of evolution.

We consider here the relative entropy~\cite{NC} of two density matrices $\rho$ and $\gamma$, defined by: 
\begin{equation}
 S(\rho\vert\vert\gamma)={\rm Tr}[\rho(\ln\rho-\ln\gamma)]
\end{equation}  
which is positive  and  vanishes if and only if $\rho\equiv \gamma$. Under CP, trace preserving dynamical maps $\Phi$,  the relative entropy 
obeys monotonicity property~\cite{MBR}  i.e., 
\begin{equation}
\label{relmap} 
 S[\Phi(\rho)\vert\vert\Phi(\gamma)]\leq S(\rho\vert\vert\gamma)  
\end{equation}  
Thus, it follows that
\begin{eqnarray} 
\label{smon}
S[\rho(t)\vert\vert\rho(t+\tau)]&\equiv& S[A(t)\rho(0)\vert\vert A(t)\rho(\tau)]\nonumber \\ 
 &\leq & S[\rho(0)\vert\vert\rho(\tau)]
\end{eqnarray} 
under a trace preserving CP map $A:\,  \rho(0)\rightarrow \rho(t)=A(t)\rho(0),$ obeying the Markovian semigroup property (\ref{MarA}).    
In other words, the {\em relative entropy difference} defined as, 
\begin{equation}
\label{relD}
S(t,\tau)=S[\rho(0)\vert\vert\rho(\tau)]-S[\rho(t)\vert\vert\rho(t+\tau)]
\end{equation}
is necessarily positive for all quantum states $\rho(t)$ evolving under CP Markovian dynamics. 
The inequality (\ref{smon}) need not be satisfied by both NCP processes as well as by a CP evolution, which departs from the semi-group property (\ref{MarA}). Thus, violation of the inequality (\ref{smon}) i.e.,
\begin{equation} 
\label{REineq}
S(t,\tau)< 0
\end{equation} 
signifies a non-Markovian  dynamical process (both  CP {\em as well as}  NCP).

We also recall here that the fidelity function~\cite{Jozsa}  defined by,  
\begin{equation}
\label{fidelity} 
F[\rho(t),\rho(t+\tau)]=\left\{{\rm Tr}\left[\sqrt{\sqrt{\rho(t)}\rho(t+\tau)\sqrt{\rho(t)}}\right]\right\}^2, 
\end{equation}
never decreases from its initial value $F[\rho(0),\rho(\tau)]$~\cite{AKRU} for the state $\rho(t)$ undergoing  a Markovian CP dynamical process.  
A sufficient criterion for non-Markovianity is therefore registered -- if the  {\em fidelity difference} function~\cite{AKRU} 
\begin{equation}
\label{fd} 
G(t, \tau)= \frac{F[\rho(t),\rho(t+\tau)] - F[\rho(0),\rho(\tau)]}{F[\rho(0),\rho(\tau)]}
\end{equation} 
assumes  {\em negative} values under open system dynamics.

The relative entropy and fidelity exhibit contrasting physical implications: The relative entropy $S[\rho(t)\vert\vert\rho(t+\tau)]$   measures the {\em instantaneous distinguishability} of the dynamical state $\rho(t+\tau)$ with its earlier time density matrix $\rho(t)$ and it declines -- when the system undergoes a CP Markovian process -- from its initial value $S[\rho(0)\vert\vert\rho(\tau)]$ to its minimum value asymptotically  i.e.,  $\lim_{t\rightarrow \infty}S(\rho(t)\vert\vert\rho(t+\tau))=0$.
On the other hand, the  fidelity $F[\rho(t), \rho(t+\tau)]$ signifies the overlap of the dynamical states $\rho(t+\tau)$,  $\rho(t)$; under any CP Markovian dynamics, it increases monotonically from its intial value $F[\rho(0),\rho(\tau)]$ to its maximum value $\lim_{t\rightarrow \infty}F[\rho(t),\rho(t+\tau)]=1$. 

The negative values of relative entropy difference (\ref{relD}) and the fidelity difference (\ref{fd}) point out that the time evolution is {\em not} a CP Markovian process --  though their positive values do not necessarily suggest that the dynamics is Markovian.  In other words, the negative values of relative entropy and fidelity differences serve as sufficient -- but not necessary -- tests of non-Markovianity (CP as well as NCP). Further, it is  not possible to draw any clear-cut inference towards whether the open-system dynamics is NCP or not --  based entirely on the negative values of the quantities (\ref{relD}), (\ref{fd}).  However, these signatures of non-Markovianity, in terms of relative entropy/fidelity, offer an operational advantage that they require only the specification of the initial density matrix  $\rho(0)$, and the dynamically evolved one $\rho(t)$ for their evaluation --  without any apriori knowledge on the nature of the  environment and/or the coupling between the system-environment.

\section{Illustrative examples}

We now proceed to investigate three different examples of two qubit initial states, jointly undergoing the unitary transformation (\ref{ue}) so that the time evolution of the first qubit (system) is represented by the dynamical map
\begin{equation}
\rho_1(0)\rightarrow \rho_1(t)=A\,\rho_1(0)={\rm Tr}_2\left[U(t)\rho_{12}(0)U(t)^\dag\right]. 
\end{equation}    
The examples are indeed expected to reveal non-Markovian features as the environment consists of just a single qubit, with no additional assumptions on  the weak  coupling limit~\cite{Breuer} invoked (so as to lead to CP Markovian dynamics, when initially uncorrelated states are considered~\cite{note2}). Further, as the dynamical evolution of the system qubit being governed by the  NCP dynamical map (\ref{Aex}),  non-Markovianity is bound to emerge.  Here, we focus on the  non-Markovian features by verifying that the relative entropy difference (\ref{relD}) and the fidelity difference (\ref{fd}) assume negative values under this open system NCP dynamics.

\subsection{Example 1: Pure entangled two qubit state}
 
 We first consider a two qubit pure entangled state
 \begin{equation} 
 \vert\Psi_{EP}\rangle=\frac{1}{\sqrt{3}}\, \left(e^{-i\phi}\vert 0_1,1_2\rangle+e^{i\phi}\vert 1_1,0_2\rangle+\vert 1_1,1_2\rangle\right),
 \end{equation}
 with the corresponding intial density matrix of the system qubit given by, 
 \begin{equation}
 \label{ex1i}
 \rho_{1}(t=0)=\frac{1}{3}\left(\begin{array}{cc} 1 & e^{-i\phi}\\ e^{i\phi} & 2 \end{array}\right)
 \end{equation}  
 in the standard basis $\{\vert 0\rangle,\vert 1\rangle\}.$

 Under the unitary transformation (\ref{ue}) we explicitly obtain the dynamical state of the system qubit as, 
 \begin{eqnarray}
 \label{ex1f}
 \rho_{1}(t)&=&{\rm Tr}_2[U(t)\vert\Psi_{EP}\rangle\langle\Psi_{EP}\vert U^\dag(t)] \nonumber \\
  &=&\frac{1}{3}\left(\begin{array}{cc} 1 & C\, e^{-i\phi}-iS\, e^{-2i\phi}\\ 
 C\, e^{i\phi}+iS\, e^{2i\phi} & 2 \end{array}\right). 
 \end{eqnarray}

This may also be identified to be the result of the action of the  open system dynamical  $A$-map (\ref{Aex}): 
\begin{widetext}
\begin{eqnarray}
\frac{1}{3}\, \left(\begin{array}{c} 1\\ C\, e^{-i\phi}-iS\, e^{-2i\phi} \\ 
 C\, e^{i\phi}+iS\, e^{2i\phi} \\ 2\end{array}  \right)=
 \frac{1}{3}\, \left(\begin{array}{cccc} 
  1 & 0 & 0 & 0 \\ \frac{1}{2}\,S\,a^* &  C & 0 & \frac{1}{2}\, S\,a^*\\  
 \frac{1}{2}\, S\, a & 0  & C & \frac{1}{2}\, S\, a \\ 
 0 & 0 & 0 & 1
 \end{array}\right)\,  \left(\begin{array}{c} 1\\ e^{-i\phi} \\ 
 e^{i\phi} \\ 2\end{array}  \right)
\end{eqnarray}
 \end{widetext}
  with the  initial state parameters (see (\ref{a1a2})) 
 \begin{eqnarray}
 a_1&=&-\langle \Psi_{EP}\vert\sigma_{1y}\sigma_{2x}\vert \Psi_{EP}\rangle=-\frac{2}{3}\sin(2\phi),\nonumber \\ 
  a_2&=&\langle \Psi_{EP}\vert\sigma_{1x}\sigma_{2x}\vert \Psi_{EP}\rangle=\frac{2}{3}\cos(2\phi), 
 \end{eqnarray} 
 governing the open system dynamics.

We obtain, after simplification, the  relative entropy $S[\rho_1(t)\vert\vert\rho_1(t+\tau)]$ of the dynamical state  (\ref{ex1f}) of the qubit as,  
\begin{widetext}
\begin{equation} 
S[\rho_1(t)\vert\vert\rho_1(t+\tau)]= \Lambda^{(+)}(t)\, \ln\left\{ \frac{\Lambda^{(+)}(t)}{[\Lambda^{(+)}(t+\tau)]^{\delta(t)}[\Lambda^{(-)}(t+\tau)]^{\nu(t)}} \right\}+\Lambda^{(-)}(t)\, \ln\left\{ \frac{\Lambda^{(-)}(t)}{[\Lambda^{(+)}(t+\tau)]^{\nu(t)}[\Lambda^{(-)}(t+\tau)]^{\delta(t)}} \right\}
\end{equation}
where we have denoted,
\begin{eqnarray}
\label{Lambda}
\Lambda^{(\pm)}(t)&=&\frac{3\pm\kappa(t)}{6},\ \  \kappa(t)=\sqrt{5-4\,\sin(2\omega\,t)\, \sin\phi}  \nonumber   \\
\delta(t)&=&\frac{1}{2\,\kappa(t)\,\kappa(t+\tau)}\left\{ \left[\kappa(t) \kappa(t+\tau) +1\right]+ 4\left\{\cos(\omega\tau)-
\sin[\omega\,(2t+\tau)]\, \sin\phi\right\}\right\}, \\ 
\nu(t)&=&\frac{1}{2\kappa(t)\,\kappa(t+\tau)}\left\{[\kappa(t)+1][\kappa(t+\tau)-1]-4\left\{\cos(\omega\tau)-
\sin[\omega\,(2t+\tau)]\, \sin\phi\right\}\right\}. \nonumber
\end{eqnarray}
\end{widetext}

In order to compute the fidelity $F[\rho_1(t),\rho_1(t+\tau)]$,  we make use of its simplified form in the case of single qubit states~\cite{Jozsa}:
\begin{eqnarray}
 F[\rho_1(t),\rho_1(t+\tau)]={\rm Tr}[\rho_1(t)\,\rho_1(t+\tau)]\nonumber \\ 
 \ \ \  +2\, \sqrt{\det\rho_1(t)\, \det\rho_1(t+\tau)}.
\end{eqnarray} 
We obtain the fidelity $F[\rho_{1}(t),\rho_{1}(t+\tau)]$ of the dynamical state (\ref{ex1f}) as,    
\begin{widetext}
\begin{equation}
F[\rho_{1}(t),\rho_{1}(t+\tau)]=
\frac{1}{9}\left\{5+2\cos(\omega \tau)-2\sin[\omega(2t+\tau)]\sin \phi
+2\sqrt{[1+\sin (2 \omega t) \,\sin \phi)( 1+\sin [2 \omega (t+\tau)]\sin \phi)}\right\}.
\end{equation}
\end{widetext}
\begin{figure}
\includegraphics*[width=2.2in,keepaspectratio]{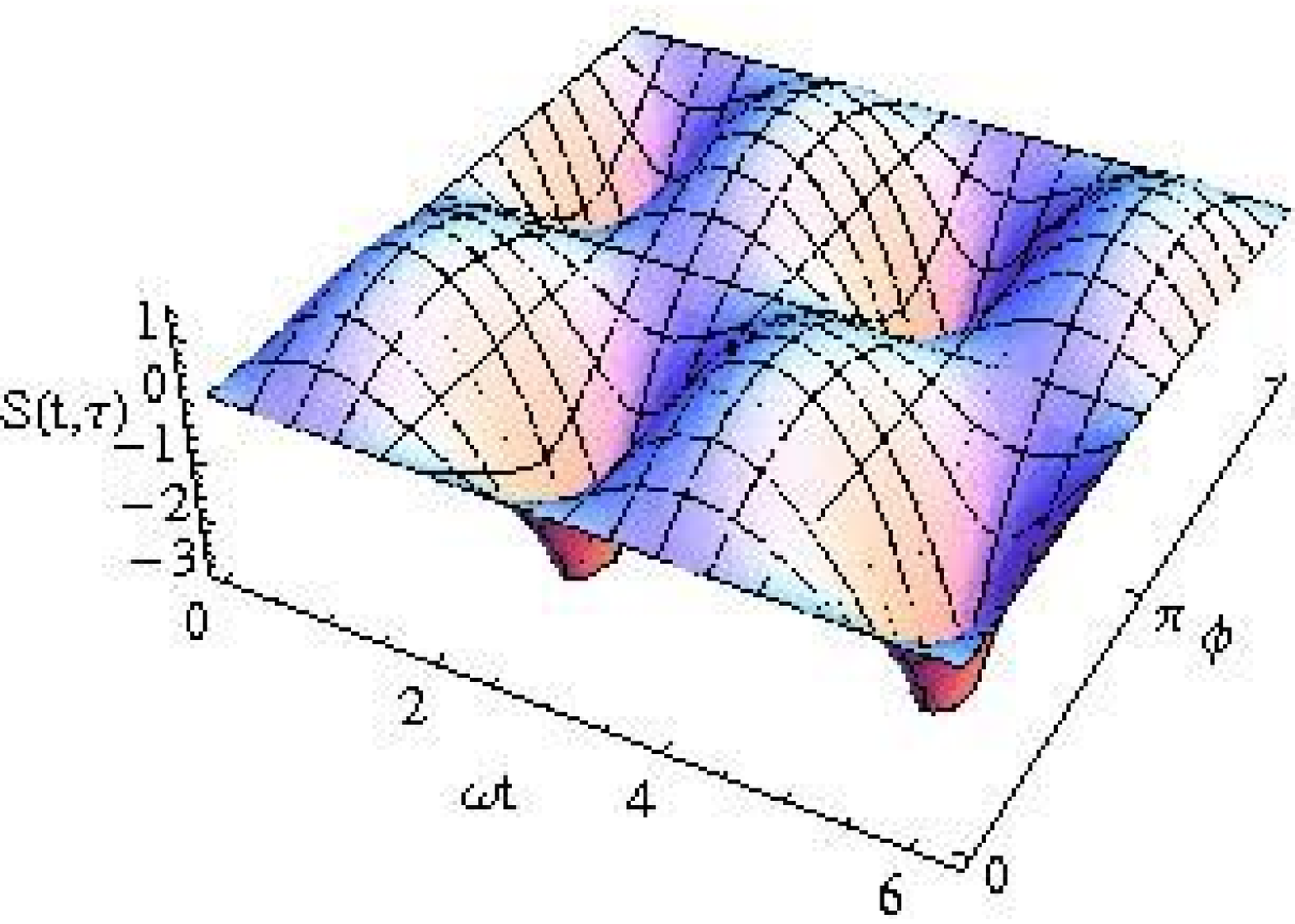}
\includegraphics*[width=2.2in,keepaspectratio]{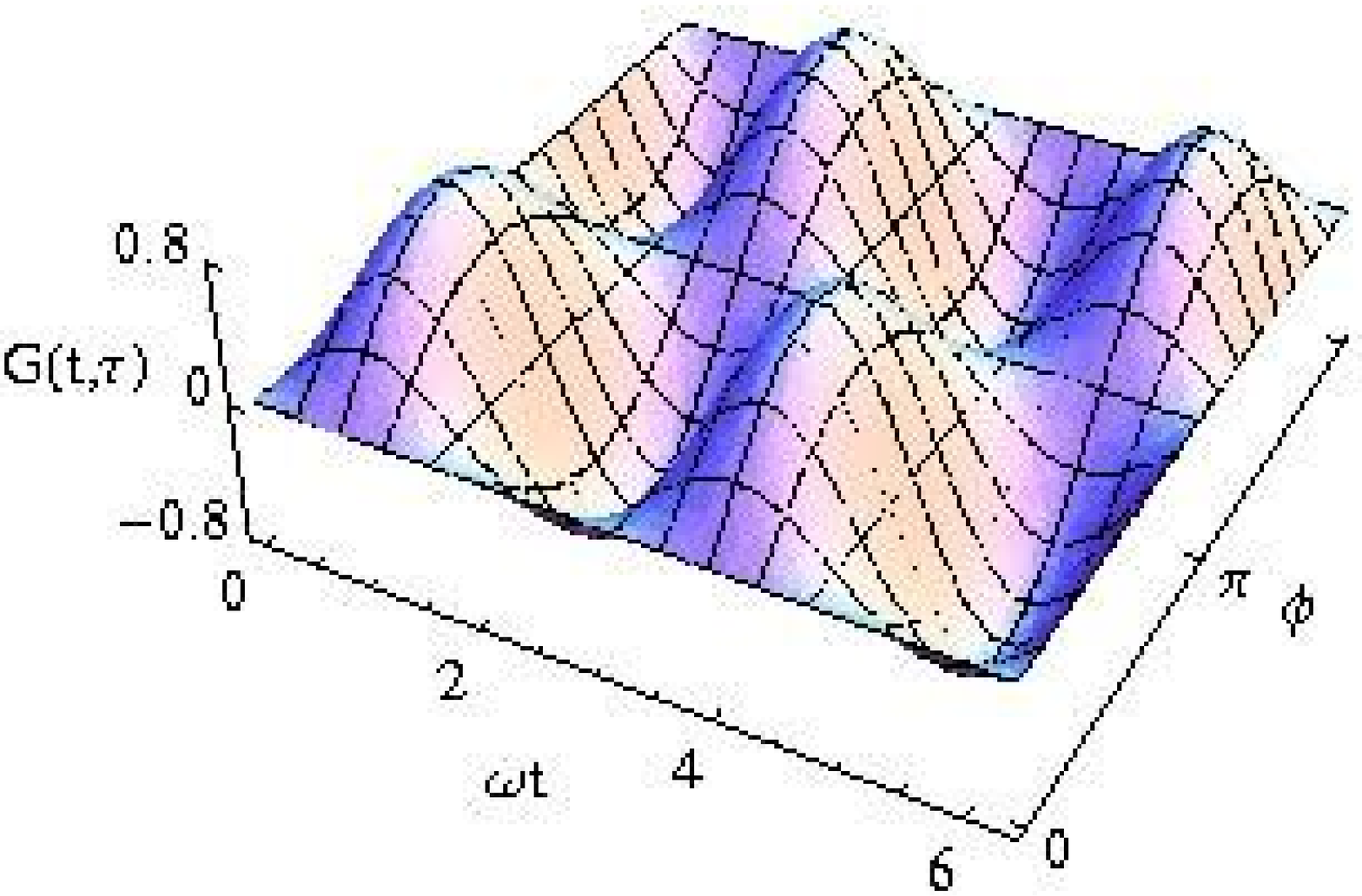}
\caption{(Color online). The relative entropy difference $S(t,\tau)$ and the fidelity difference $G(t, \tau)$ corresponding to the dynamical state (\ref{ex1f}), as a function of  $\omega\,t$ and $\phi$;  $\omega\tau=\pi$. Negative regions of $S(t,\tau)$, $G(t,\tau)$ point towards non-Markovian behavior. All quantities are dimensionless.}
\end{figure}

The relative entropy difference $S(t,\tau)$ and the fidelity difference $G(t,\tau)$ (see (\ref{fd})) for the dynamical state (\ref{ex1f}) are plotted as a function of $\omega\, t$ and $\phi$ in Fig.~1 -- where the negative regions  of both $S(t,\tau)$, $G(t,\tau)$  display  non-Markovianity of the dynamical process. 

\subsection{Example 2: Two qubit Werner state} 

We consider  Werner state of two qubits as the initial system-environment state
\begin{equation}
\rho_{W}(t=0)=\frac{x}{4} I_1\otimes I_2+(1-x)\, \vert\Psi_-\rangle\langle\Psi_-\vert 
\end{equation}
where $\vert\Psi_-\rangle=\frac{1}{\sqrt{2}}\left(\vert 0_1, 1_2\rangle-\vert 1_1,0_2\vert\rangle\right).$ The initial state of the system qubit is given by 
$\rho_{W1}(0)~=~{\rm Tr}_2[\rho_{W}(0)]~=~\frac{1}{2}\, I_1.$ 

Under open system dynamics (\ref{ue}) we obtain the dynamical state of the system qubit as, 
\begin{eqnarray}
\label{wt}
\rho_{W1}(t)&=&{\rm Tr}_2[U(t)\rho_{W}(0)U^\dagger(t)]  \\
&=&\frac{1}{2}\left(\begin{array}{cc} 1 & -i\,(1-x)\, \sin(\omega t) \\ 
i\, (1-x)\, \sin(\omega t) & 1 \end{array} \right), \nonumber 
\end{eqnarray} 
(which may also be directly obtained by employing the  $A$-map  (\ref{Aex}) as, 
$\left[\rho_{W1}(t)\right]_{r's'}=\sum_{r,s}\, A_{r's';rs}(t)\,\left[\rho_{W1}(0)\right]_{rs}$ with $a_1=-{\rm Tr}[\rho_{W}(0)\,\sigma_1y\sigma_{2x}]=0, 
a_2={\rm Tr}[\rho_{W}(0)\,\sigma_{1x}\sigma_{2x}]=(1-x)$).

The relative entropy $S[\rho_{W1}(t)\vert\vert\rho_{W1}(t+\tau)]$  of the system qubit (\ref{wt}) is identified to be, 
\begin{eqnarray}
 &&S[\rho_{W1}(t)\vert\vert\rho_{W1}(t+\tau)]=  p_{+}(t)\, \ln\left[\frac{p_{+}(t)}{p_{+}(t+\tau)}\right]\nonumber \\ 
 &&\hskip 0.9in  +p_{-}(t)\, \ln\left[\frac{p_{-}(t)}{p_{-}(t+\tau)}\right],
\end{eqnarray} 
 where 
 \begin{equation} 
 p_{\pm}(t)=\frac{1}{2}[1\pm (1-x)\, \sin(\omega t)].
 \end{equation}
Further, the fidelity $F[\rho_{W1}(t),\rho_{W1}(t+\tau)]$ is obtained as, 
\begin{eqnarray}
&& F[\rho_{W1}(t),\rho_{W1}(t+\tau)]= p_{+}(t)\, p_{+}(t+\tau) 
 +p_{-}(t)\, p_{-}(t+\tau) \nonumber \\ 
&& \hskip 0.6in +2\,\sqrt{p_{+}(t)p_{-}(t)p_{+}(t+\tau)p_{-}(t+\tau)} 
\end{eqnarray}
\begin{figure}
\includegraphics*[width=2.2in,keepaspectratio]{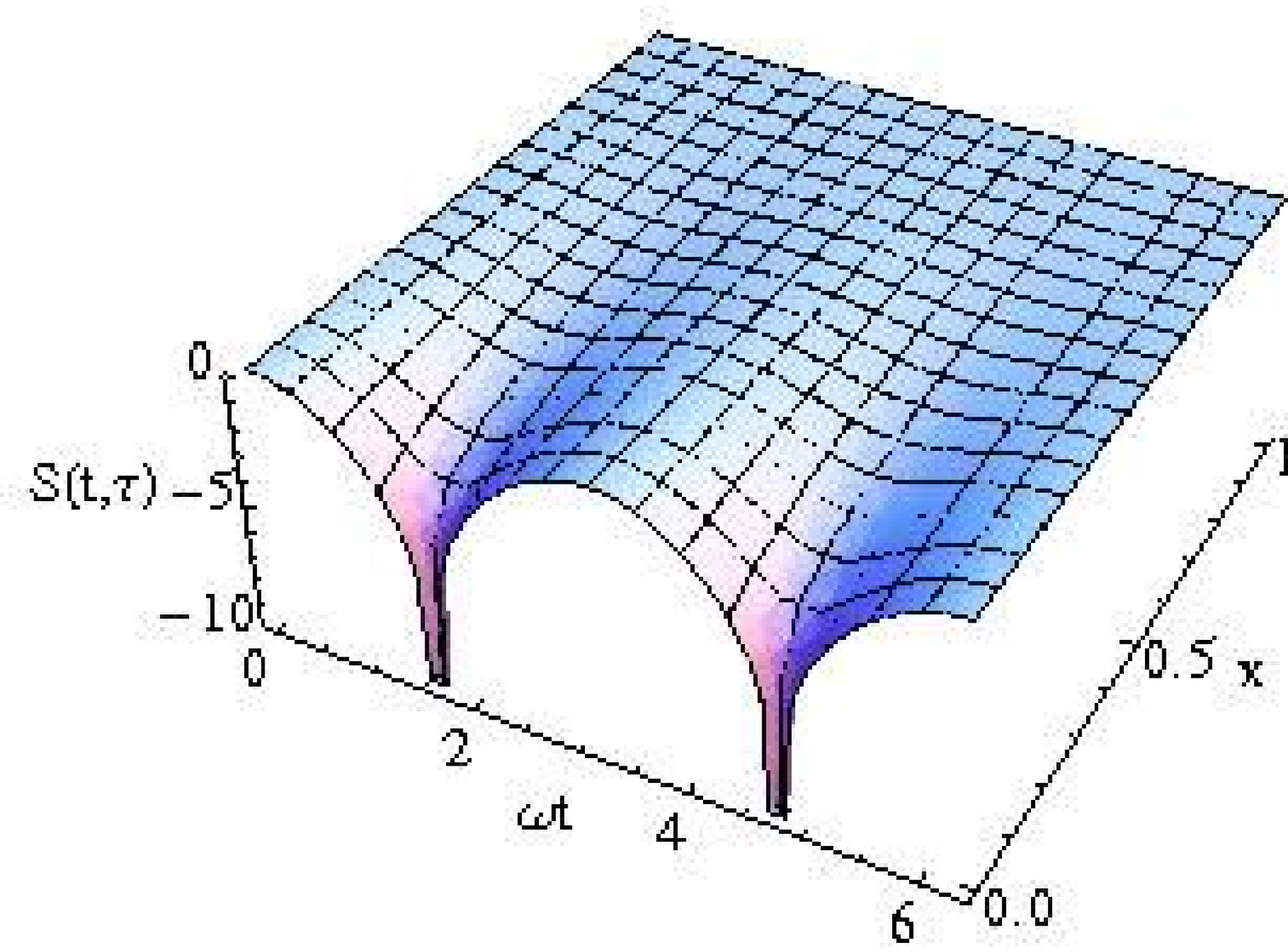}
\includegraphics*[width=2.2in,keepaspectratio]{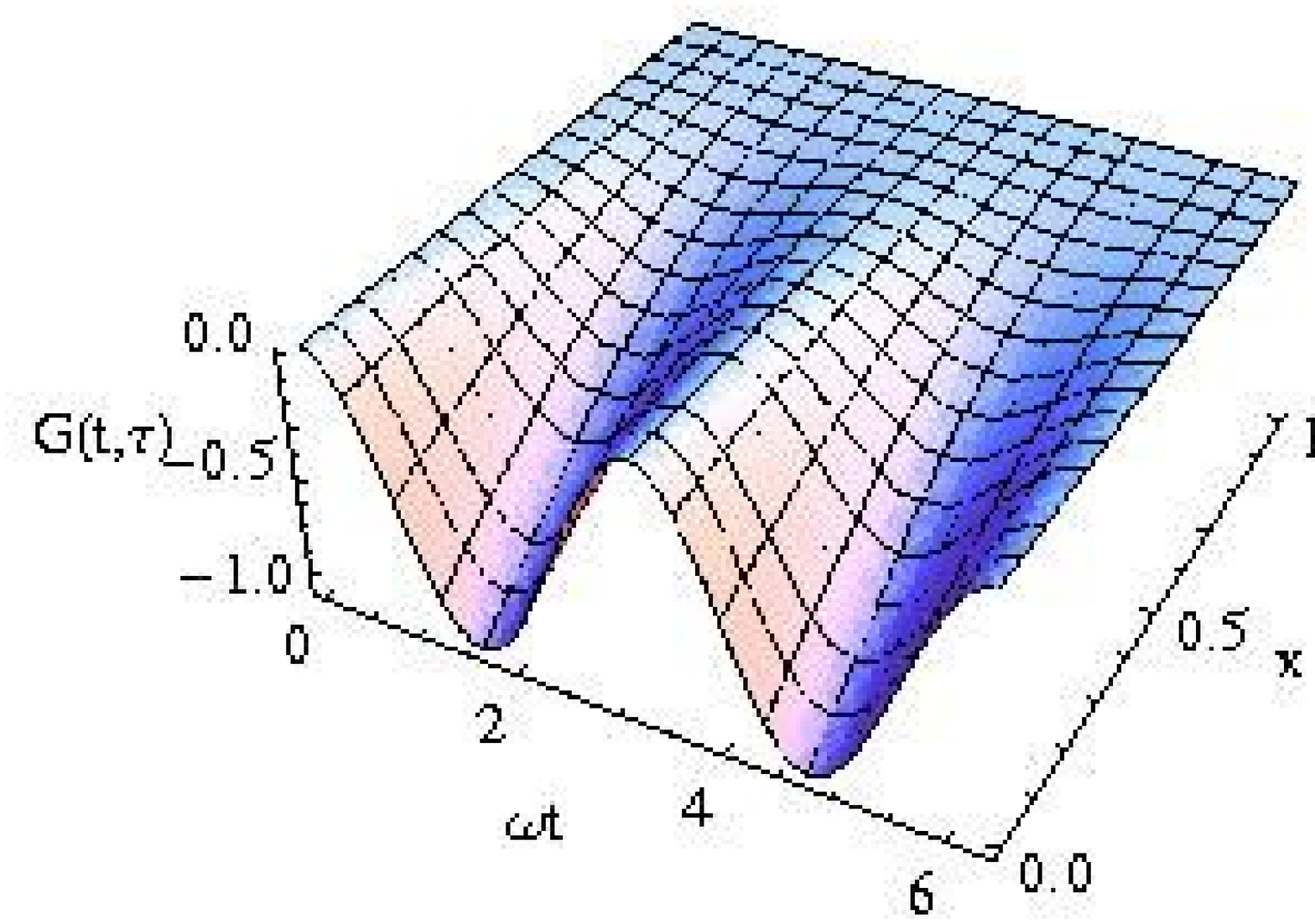}
\caption{(Color online). The relative entropy difference $S(t,\tau)$ and the fidelity difference $G(t, \tau)$ of the dynamical state (\ref{wt}), as a function of (dimensionless) time $\omega\,t$ and $x$; here, we have chosen $\omega\tau=\pi$. The functions $S(t,\tau)$, $G(t,\tau)$ are negative in almost the entire region (except for $x=1$), reflecting the non-Markovianity of the underlying NCP dynamics. All quantities are dimensionless.}
\end{figure}   

We have plotted, in Fig.~2,  the relative entropy difference $S(t,\tau)$ and the fidelity difference $G(t,\tau)$ for this dynamical example. Here too, the non-Markovianity of the dynamics is clearly depicted by the negative values of $S(t,\tau)$ and $G(t,\tau)$. 

\bigskip 

\subsection{Example 3: Mixed separable state of two qubits}  
Now we consider a separable mixed state of two qubits, 
\begin{equation}
\rho_{S}(t=0)=\frac{1}{4}\left(I\otimes I+ s_x\, \sigma_{1x}+s_y\, \sigma_{1y} +s_z\, \sigma_{1z}+\, d\, \sigma_{1y}\,\sigma_{2x}\right). 
\end{equation}    
 The initial state of the first qubit (system) is obtained to be,
\begin{eqnarray}
\rho_{S1}(0)&=&{\rm Tr}_2[\rho_{S}(0)]\nonumber \\ 
&=&\frac{1}{2}\left(\begin{array}{cc} 1+s_z & s_x-is_y \\ 
s_x+is_y & 1-s_z \end{array} \right)
\end{eqnarray}

Open system dynamics of the system qubit (corresponding to the joint unitary evolution (\ref{ue}) leads to the dynamical state $\rho_{S1}(t)$ 
  as, 
\begin{eqnarray}
\label{st}
\rho_{S1}(t)&=&{\rm Tr}_2[U(t)\rho_{S}U^\dagger(t)]  \\
&=&\frac{1}{2}
\left(\begin{array}{cc} 1+s_z & s(t) \\ 
s^*(t) & 1-s_z \end{array} \right), \nonumber 
\end{eqnarray} 
where we have denoted 
\begin{equation}
s(t)=(s_x-is_y)\cos(\omega t)-d\, \sin(\omega\, t).
\end{equation}
Note that in this example, we have the intial dynamical parameters $a_1=-{\rm Tr}[\rho_{S}(0)\,\sigma_1y\sigma_{2x}]=-d, 
a_2={\rm Tr}[\rho_{S}(0)\,\sigma_{1x}\sigma_{2x}]=0$ and the dynamical state of the system  (\ref{st})  
is equivalently obtained by transforming the initial density matrix (expressed as a column) through the $A$ matrix (\ref{Aex}).

We evaluate the relative entropy $S[\rho_{S1}(t)\vert\vert\rho_{S1}(t+\tau)]$  of the state (\ref{st}) to obtain,   
\begin{widetext}
\begin{eqnarray}
 &&S[\rho_{S1}(t)\vert\vert\rho_{S1}(t+\tau)]=\Omega^{(+)}(t)\, \ln\left\{ \frac{\Omega^{(+)}(t)}{[\Omega^{(+)}(t+\tau)]^{\mu(t)}[\Omega^{(-)}(t+\tau)]^{\eta(t)}} \right\}+
\Omega^{(-)}(t)\, \ln\left\{\frac{\Omega^{(-)}(t)}{[\Omega^{(+)}(t+\tau)]^{\eta(t)}[\Omega^{(-)}(t+\tau)]^{\mu(t)}}\right\}, \nonumber \\
\end{eqnarray} 
where
\begin{eqnarray}
\label{zeta}
\Omega^{(\pm)}(t)&=&\frac{1}{2}[1\pm \zeta(t)], \ \  
\zeta(t)=\sqrt{s_z^2+\chi(t)}, \ \ \chi(t) =\left[s_x \cos(\omega\,t) 
-d \sin(\omega\,t)\right]^2+s_y^2 \cos^2(\omega\,t) \nonumber \\ 
\mu(t)&=&\frac{1}{4\zeta(t)\,\zeta(t+\tau)}\left\{\frac{\chi(t) \chi(t+\tau)}{[\zeta(t)-s_z][\zeta(t+\tau)-s_z]}+[\zeta(t)-s_z]
[\zeta(t+\tau)
-s_z]+2\, {\cal R}(t)\right\} \\
\eta(t)&=&\frac{1}{4\zeta(t)\zeta(t+\tau)} \left\{ \frac{\chi(t)\,[\zeta(t+\tau)-s_z]}{\zeta(t)-s_z}+
\frac{ \chi(t+\tau)\,[\zeta(t)-s_z]}{\zeta(t+\tau)-s_z}-2\,{\cal R}(t)  \right\}\nonumber  \\ 
\label{R}
{\cal R}(t)&=&\left(s_x^2+s_y^2\right)\cos(\omega\,t) \cos[\omega\,(t+\tau)]+d^2 \sin(\omega\, t) \sin[ \omega\, (t+\tau)]- d\,s_x \sin [\omega\, (2t+\tau)]. \nonumber
\end{eqnarray}
The fidelity $F[\rho_{S1}(t),\rho_{S1}(t+\tau)]$ associated with the dynamical state (\ref{st}) is found to be,  
\begin{equation}
 F[\rho_{S1}(t),\rho_{S1}(t+\tau)]=\frac{1}{2}\left[ 1+s_z^2+\sqrt{[1-\zeta^2(t)][1-\zeta^2(t+\tau)]}+{\cal R}(t) \right],
\end{equation}
where $\zeta(t),\ {\cal R}(t)$ are defined in (\ref{zeta}). 
\begin{figure}
\includegraphics*[width=2.2in,keepaspectratio]{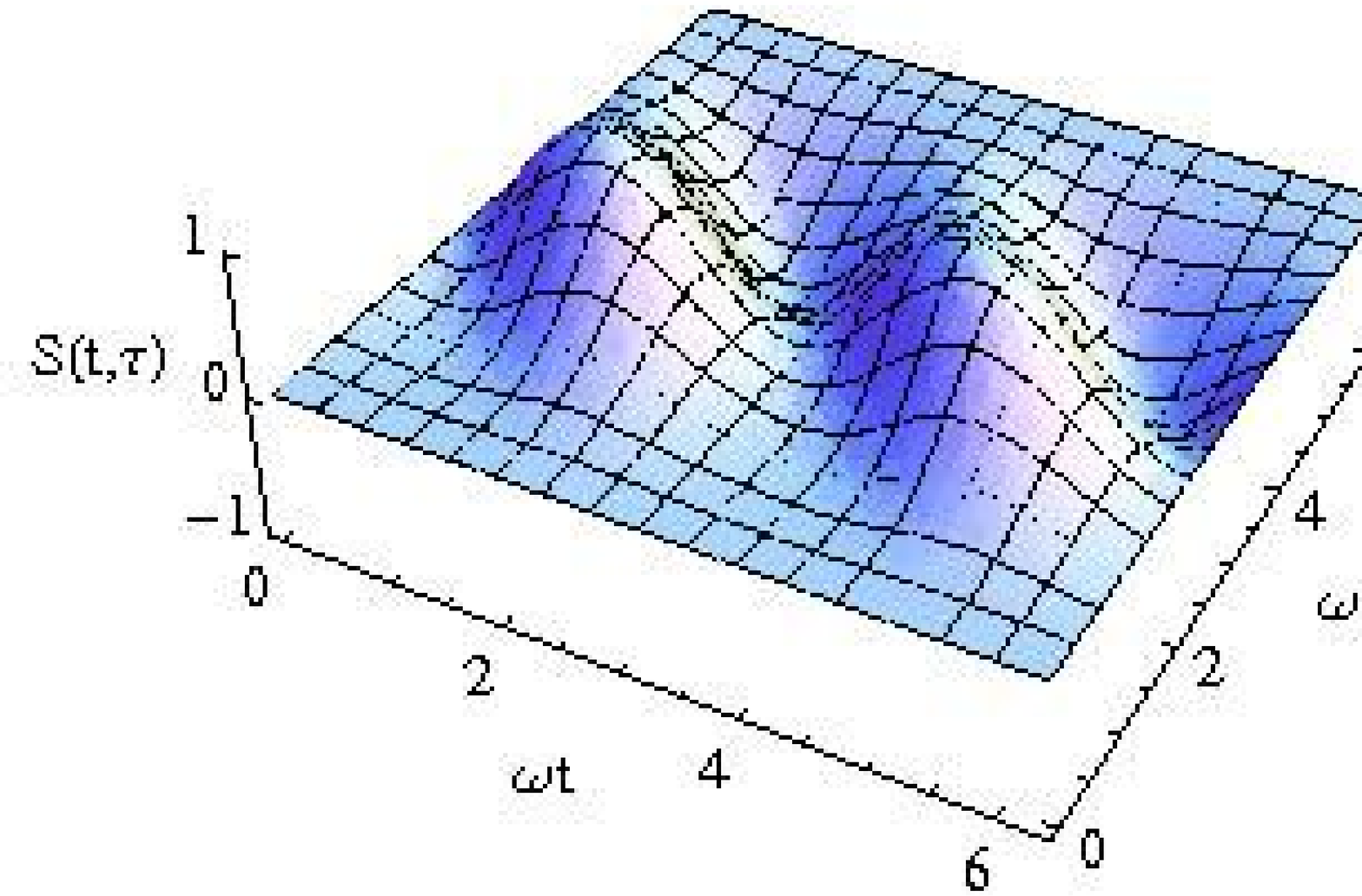}
\includegraphics*[width=2.2in,keepaspectratio]{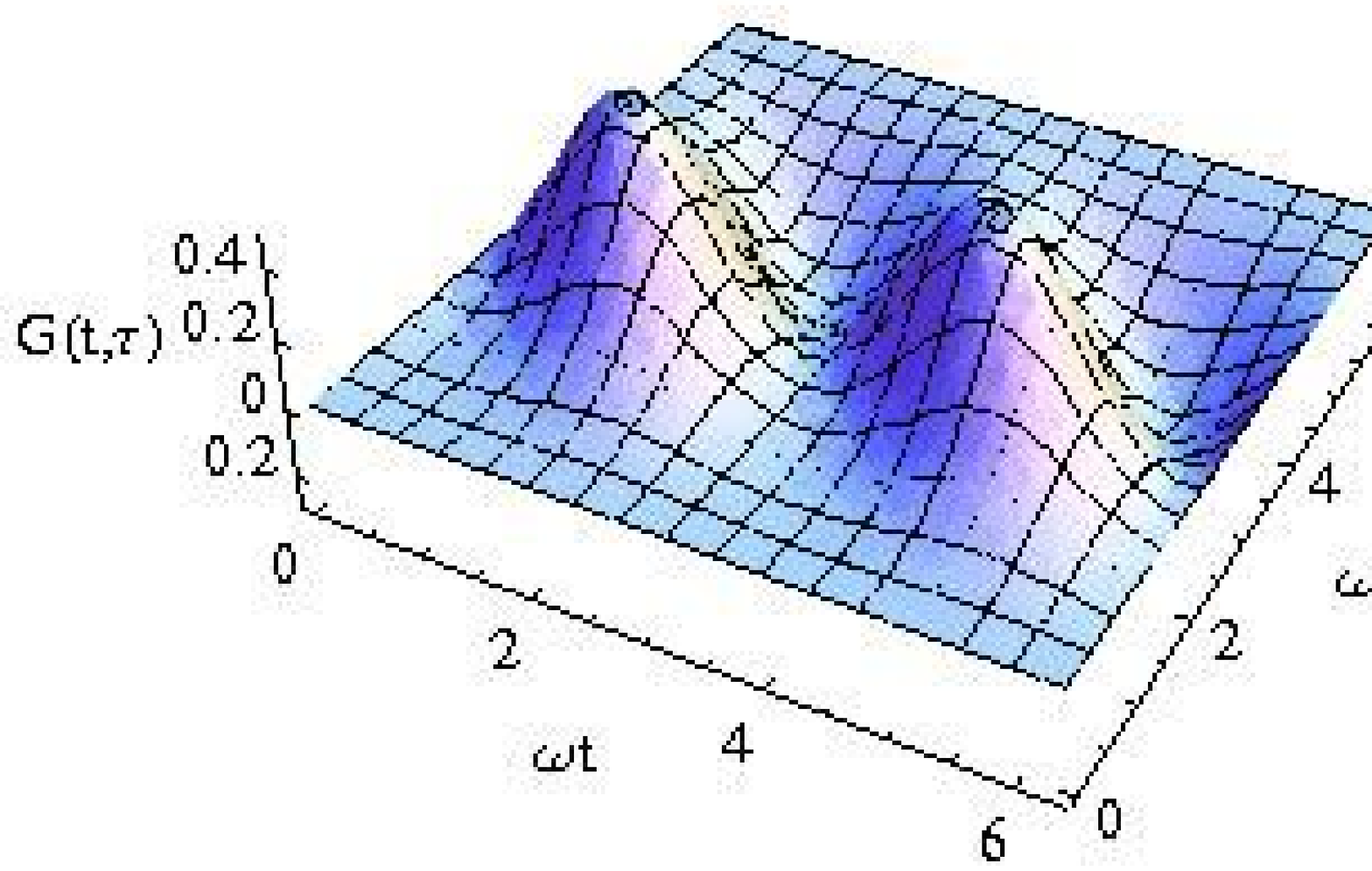}
\caption{(Color online). The relative entropy difference $S(t,\tau)$ and the fidelity difference $G(t, \tau)$ of the dynamical state (\ref{st}), as a function of (dimensionless) time $\omega\,t$ and $\omega\tau$; here, we have chosen $s_x=s_y=s_z=d=1/\sqrt{6}$. Negative fluctuations of the functions $S(t,\tau)$, $G(t,\tau)$ reveal non-Markovianity. All quantities are dimensionless.}
\end{figure}   
\end{widetext}
In Fig.~3, we depict the relative entropy difference $S(t,\tau)$ and the fidelity difference $G(t,\tau)$ of the state (\ref{st}). The negative values of $S(t,\tau)$ and $G(t,\tau)$ demonstrate the non-Markovianity of the dynamical process in this example too.  

We emphasize here that the negative values of the relative entropy difference and fidelity difference serve as {\em qualitative}  reflections of deviation from a CP Markovian behavior -- with no specific importance attached to the degree of negativity.  Further, the positive regions in the figures, {\em do not} indicate that the process is being Markovian in these regions -- because our criteria offer only sufficient (not necessary) tests of non-Markovianity.  It is also worth pointing out here that had we considered  initially uncorrelated (or zero discord) states,  a CP dynamical map associated with the same unitary dynamics (\ref{ue}), would also be expected to exhibit non-Markovianity~\cite{note2}. In other words, it is not evident if the departure from Markovianity is entirely from the NCP nature (resulting due to intially correlated states) or from the built-in non-Markovianity associated with the dynamics. It remains  an open question to recognize distinct signatures of  NCP non-Markovianity emerging exclusively due to initial system-environment correlations.

\section{Summary}
 
While open system evolution was formulated nearly five decades ago -- with the introduction of dynamical $A$ and $B$ maps --  by Sudarshan et. al~\cite{ECGS1, Jordan1}, several interesting questions on the nature of dynamical maps are being raised recently~\cite{Jordan,Rosario,ECGS,Modi,Pechukas, Alicki,Buzek,Terno,Lidar}.  It has been recognized that NCP dynamical maps make their presence felt in the reduced dynamics obtained from the joint unitary evolution, if the system and environment are  in an initially correlated state~\cite{Jordan,Rosario,ECGS,Modi}. Conceptual understanding of positive -- but NCP dynamical maps -- has thus been attracting increasing attention. 
In this  paper, we have  developed a canonical structure for the $A$-map, and shown  that this canonical $A$-map offers an elegant approach to investigate if the dynamics is  CP/NCP. Manifestations of memory effects in CP open system quantum evolution has been investigated in a previous work by some of us~\cite{AKRU} and here, we have focused on exploring departure of CP Markovianity in a specifically chosen NCP dynamics~\cite{Jordan} with initially correlated  system-environment states. We have proposed a test to verify deviations from CP Markovianity, based on the relative entropy of the dynamical state of the system -- which together with an analogous characterization~\cite{AKRU}, in terms of the fidelity --  is employed here to study prevalent memory effects in the NCP  evolution.  We have examined three different examples  with diverse kinds of initial correlations -- such as two qubits in (i) a pure entangled state, (ii)  Werner state ( mixed two qubit state which encompasses both entangled and separable states depending on a single real parameter $x$)   and (iii) a separable state.  All the three dynamical examples considered here display non-Markovianity. However, it is not evident if  the reflections of  non-Markovianity are essentially arising due to NCP nature of the process (initially correlated states of qubits). This leaves open an important question:   Are there any distinct signatures of non-Markovianity emerging entirely because of the NCP nature of the dynamical process?        

\section*{Acknowledgement} Sudha gratefully acknowledges the support of D.C.Pavate Foundation for the award of 
Pavate Memorial Visiting Fellowship.  She is also thankful to the local hospitality and facilities provided at 
Sidney Sussex College, Cambridge, UK.


\begin{thebibliography}{0}
\bibitem{Breuer} H.-P. Breuer and F. Petruccione, {\em The Theory of Open
Quantum Systems} (Oxford Univ. Press, Oxford, 2007).
\bibitem{Kraus} K. Kraus, States, Effects and Operations: Fundamental
Notions of Quantum Theory, vol. 190 of Lecture notes in Physics (Spring-Verlag, New York, 1983). 
\bibitem{ECGS1} E. C. G. Sudarshan, P. Mathews, and J. Rau, Phys. Rev. {\bf 121}, 920 (1961).
\bibitem{Jordan1} T. F. Jordan and E. C. G. Sudarshan, J. Math. Phys. {\bf 2}, 772 (1961).
\bibitem{Jordan} T. F. Jordan, A. Shaji, and E. C. G. Sudarshan, \pra {\bf 70}, 052110 (2004).
\bibitem{Rosario} C. A. Rodr{\' i}guez-Rosario, K. Modi,  Aik-meng Kuah, A. Shaji, and E. C. G. Sudarshan, J. Phys. A {\bf 41}, 205301 (2008). 
\bibitem{ECGS} C. A. Rodr{\' i}guez-Rosario and E. C. G. Sudarshan, arXiv:0803.1183 (quant-ph).
\bibitem{Modi} K. Modi and E. C. G. Sudarshan, \pra {\bf 81}, 052119 (2010).  
\bibitem{Pechukas} P. Pechukas, \prl {\bf 73}, 1060 (1994).
\bibitem{Alicki} R. Alicki, \prl {\bf 75}, 3020 (1995); P. Pechukas, \prl {\bf 75}, 3021 (1995). 
\bibitem{Buzek} P. Stelmachovic and V. Buzek, \pra {\bf 64}, 062106(2001).
\bibitem{Terno} H. A. Carteret, D. R. Terno, and K. {\. Z}yczkowski, \pra {\bf 77}, 042113 (2008). 
\bibitem{Lidar} A. Shabani and D. A, Lidar, \prl {\bf 102}, 100402 (2009).
\bibitem{Zurek} H. Ollivier and W. H. Zurek, \prl {\bf 88}, 017901 (2001).
\bibitem{Lindblad} G. Lindblad, Comm. Math. Phy. {\bf 48}, 119 (1976). 
\bibitem{GKS} V. Gorini, A. Kossakowski, and E. C. G. Sudarshan, J. Math. Phys. {\bf 17}, 821 (1976).
 \bibitem{Daffer} S. Daffer, K. W{\' o}dkiewicz, J. D. Cresser, and J. K. McIver \pra {\bf 70}, 010304 (2004).
\bibitem{Breuer2}  H.-P. Breuer, \pra {\bf 70}, 012106 (2004).
\bibitem{Kossakowski} D. Chrus{\' c}in{\' s}ki, and A. Kossakowski, \prl {\bf 104}, 070406 (2010); D. Chrus{\' c}in{\'s}ki, and A. Kossakowski and 
S. Pascazio, \pra {\bf 81}, 032101 (2010).
\bibitem{Cirac} M.M. Wolf, J. Eisert, T.S. Cubitt, and J.I. Cirac, \prl {\bf 101}, 150402 (2008).
\bibitem{B3} H.-P. Breuer et al. , E.-M. Laine, J. Piilo, Phys. Rev.
Lett. 103, 210401 (2009); E.-M. Laine, J. Piilo, H.-P. Breuer, \pra {\bf 81}, 062115 (2010).
\bibitem{Angel} A. Rivas, S. F. Hulega, M. B. Plenio, \prl {\bf 105}, 050403 (2010).
\bibitem{AKRU} A. K. Rajagopal, A. R. Usha Devi, R. W. Rendell, \pra {\bf 82}, 042107 (2010).  
\bibitem{Choi} M. D. Choi, Can. J. Math. {\bf 24}, 520 (1972); Linear Algebra and Appl. {\bf 10}, 285 (1975).  
\bibitem{note} The eigenvalues of ${\cal A}$ are the same as that of the $B$-matrix considered in Ref.~\cite{Jordan}.  
\bibitem{note2} The eigenvalues (\ref{lam4}) reduce to $\{1+\cos(\omega\, t), 0,\,1-\cos(\omega\, t),\, 0\}$ when $a_1=a_2=0$ (which corresponds to an initially uncorrelated state) and the associated dynamical map is CP. It may be readily verified -- based on the  time dependence of the associated Kraus operators (see ~\cite{AKRU}) of this CP dynamical map with uncorrelated initial state -- that the dynamics is non-Markovian.   
\bibitem{NC} M. A. Nielsen and I. L. Chaung, {\em Quantum computation and quantum information} (Cambridge Univ. Press, Cambridge, 2002).  
\bibitem{MBR} M. B. Ruskai, J. Math. Phys. {\bf 43}, 4358 (2002).
\bibitem{Jozsa} R. Jozsa, J. Mod. Optics, 41, 2315 (1994).
\end{thebibliography}
\end{document}